\documentclass[12pt,titlepage]{article}
\usepackage[letterpaper,top=1in,bottom=1in,left=1.25in,right=1.25in]{geometry}
\usepackage{graphicx,cite,url}

\usepackage{amsmath,mathrsfs,cleveref}

\begin{document}

\title{Deep speckle correlation: a deep learning approach towards scalable imaging through scattering media}

\author{Yunzhe Li$^{1}$, Yujia Xue$^{1}$, Lei Tian$^{1,*}$
\\
\multicolumn{1}{p{\textwidth}}{\centering\emph{\normalsize 1. Department of Electrical and Computer Engineering, Boston University, Boston, MA 02215, USA\\
		$^{*}$ leitian@bu.edu
}}}

\maketitle

\begin{abstract}
Imaging through scattering is an important, yet challenging problem.  
Tremendous progress  has been made by exploiting the deterministic input-output `transmission matrix' for a fixed medium.  
However, this `one-to-one' mapping is highly susceptible to speckle decorrelations -- small perturbations to the scattering medium lead to model errors and severe degradation of the imaging performance. 
  Our goal here is to develop a new framework that is highly scalable to both medium perturbations and measurement requirement.   
  To do so, we propose a statistical `one-to-all' deep learning technique that encapsulates a wide range of statistical variations for the model to be resilient to speckle decorrelations.  
  Specifically, we develop a convolutional neural network (CNN) that is able to learn the statistical information contained in the speckle intensity patterns captured on a set of diffusers having the same macroscopic parameter.
  We then show for the first time, to the best of our knowledge, that the trained CNN is able to generalize and make high-quality object predictions through an entirely different set of  diffusers of the same class.  
 Our work paves the way to a highly scalable deep learning approach for imaging through scattering media. 
\end{abstract}


\section{Introduction}

\begin{figure*}[t]
\centering\includegraphics[width=1\linewidth]{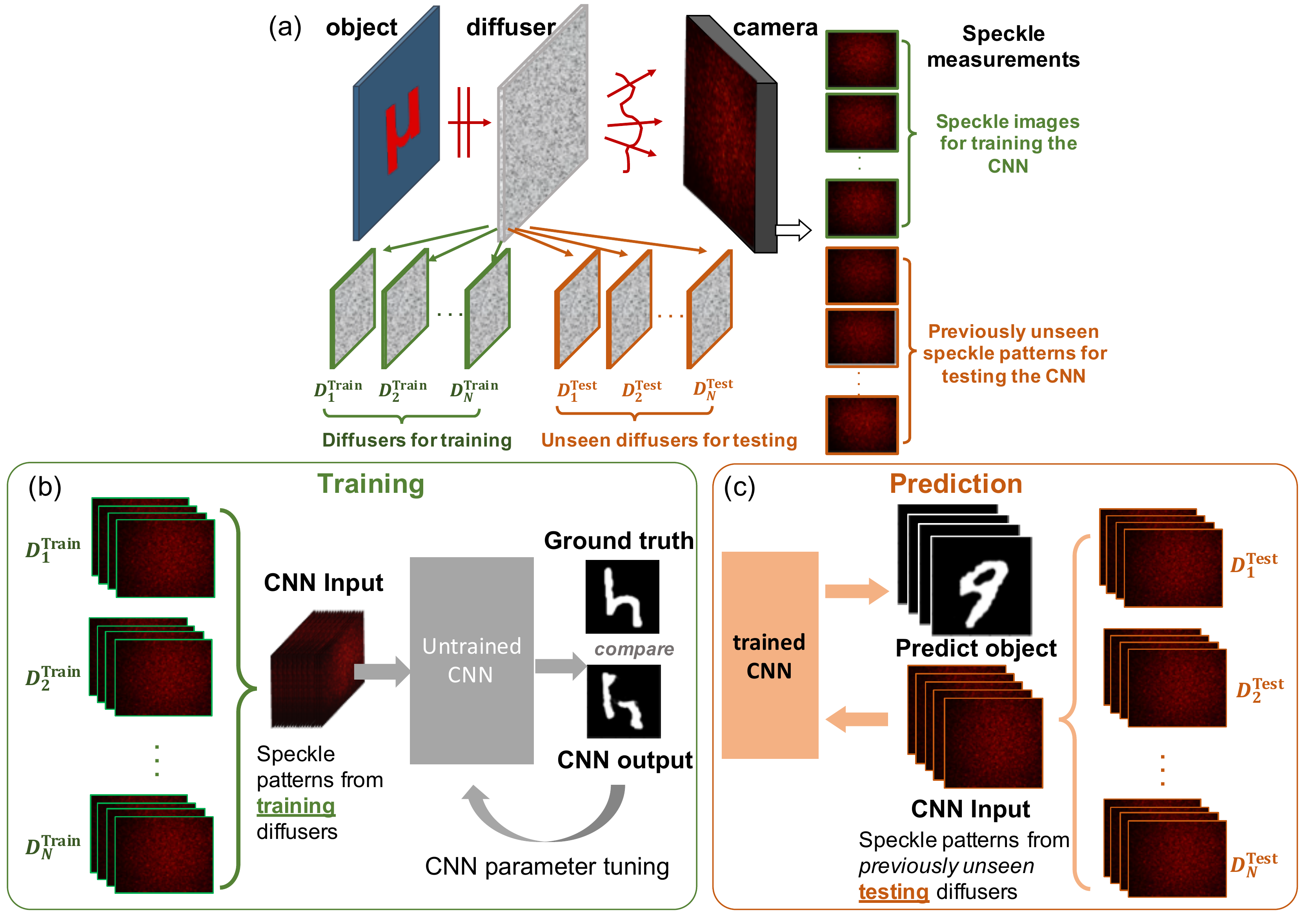}
\caption{An overview of our deep learning based imaging through scattering technique. 
(a) Speckle measurements are repeated on multiple diffusers. 
(b) During the training stage, only speckle patterns collected through the {\it training} diffusers $D^{\mathrm{train}}_1,D^{\mathrm{train}}_2,\cdots, D^{\mathrm{train}}_N$ are used. 
(c) During the testing stage, objects are predicted from speckle patterns collected through {\it previously unseen testing} diffusers $D^{\mathrm{test}}_1,D^{\mathrm{test}}_2,\cdots, D^{\mathrm{test}}_N$, demonstrating the superior scalability of our deep learning approach.}
\label{overview}
\end{figure*}

Light scattering in complex media is a pervasive problem across many areas, such as deep tissue imaging~\cite{Ntziachristos2010}, imaging in degraded environment~\cite{roggemann2018imaging}, and wavefront shaping~\cite{Vellekoop2007,mosk2012controlling,rotter2017light}.  
To date, there is no simple solution for inverting scattering because of the many possible optical paths between the object and the detector.  
The output of a coherent light scattered from a complex medium exhibits a seemingly random speckle pattern~\cite{Goodman2007}.  
The speckle's spatial distribution is a complex function of both the microscopic arrangement of the scatterers and the wavefront of the incident field.  
Thus, a comprehensive {\it deterministic} characterization of the scattering process is often difficult, requiring large-scale measurements.  

Major progress has been made by using the transmission matrix~(TM) framework~\cite{Vellekoop2007,Popoff2010a,Kim2015} that characterizes the `one-to-one' input-output relation of a fixed scattering medium as a linear shift-{\it variant} matrix. 
Due to the many underlying degrees of freedom, the TM is inevitably large, whose size generally grows quadratically as the transferred pixel number, i.e.~the system's space-bandwidth-product (SBP).  
This makes this approach highly measurement and data-demanding for high-SBP applications. 
Under special conditions, simplification can be made using the memory effect~\cite{Freund1990}, which approximates the system to be shift-{\it invariant}.  However, the SBP of this method is still small due to the limited memory effect range~\cite{Freund1990,schott2015characterization}, finite sensor dynamic range~\cite{Katz2014},  imaging geometry~\cite{tokovinin2000isoplanatism,mertz2015field,li2015conjugate}, and trade-offs between illumination coherence, speckle contrast, and measurement requirement~\cite{labeyrie1970attainment,Bertolotti2012,Katz2014,Edrei2016}.

A major limitation of these existing approaches is their high susceptibility to model errors.  The phase-sensitive TM is inherently intolerant to speckle decorrelations~\cite{hillman2013digital,jang2015relation,liu2015optical,qureshi2017vivo}. 
Slight changes of the medium can lead to much reduced correlations between the speckles measured before and after.
This indicates the breakdown of the previous input-output relation, and results in rapid degradation of the transferred images.  
In other words, a new TM is needed once the speckle patterns become decorrelated, e.g. Pearson correlation coefficient (PCC) $<1/e$,  making these methods challenging to scale for applications involving dynamic scatterers. 
Current solutions focus on developing hardware with higher speed than the medium's decorrelation time~\cite{Conkey2012,wang2015focusing,liu2015optical,liu2017focusing,blochet2017focusing}; still, they are often limited by the memory effect.

Our goal here is to develop a highly {\it scalable} imaging through scattering framework by overcoming the existing limitations in susceptibility to speckle decorrelation and SBP. 
The main approach is to build a 'one-to-all' model that possesses two essential {\it statistical} properties.  
First, `one' model sufficiently encompasses the statistical {\it variations} across `all' scattering media with different scatterer microstructures but within the same class.  
Second, the model can distill the statistically {\it invariant} information encoded in the speckle patterns (correlated or decorrelated). 
Together, they allow the single model to be generalizable to various objects/media having the same statistical characteristics.

The proposed model is built on a deep learning (DL) framework. To satisfy the desired statistical properties, we do {\it not} train a convolutional neural network (CNN) to learn the TM of {\it a single} scattering medium.  Instead, we build {\it a CNN to learn a `one-to-all' mapping by training on multiple scattering media with different microstructures while having the same macroscopic parameter.}  
Specifically, we show that our CNN model trained on a few diffusers can sufficiently support the statistical information of all diffusers having the same mean characteristics (e.g. `grits'~\cite{grits}).  We then experimentally demonstrate that the CNN is able to `invert' speckles captured from entirely different diffusers to make high-quality object predictions, as outlined in Fig.~\ref{overview}.  

DL is shown to be powerful in solving complex imaging problems, providing state-of-the-art performance in super-resolution~\cite{rivenson2017deep,wang2018deep}, holography~\cite{rivenson2018phase,ren2018learning},  and phase recovery~\cite{sinha2017lensless,nguyen2018convolutional}. 
Instead of building an explicit model, DL takes a data-driven approach that seeks solutions by learning from large-scale dataset. 
The major benefit includes the flexibility and adaptability in solving complex problems, in which a parametric model is hard to derive and/or prone to errors. 
Closely related to our work are the learning-based techniques for imaging/focusing through diffusers~\cite{horisaki2016learning,lyu2017exploit,horisaki2017learning,li2018imaging,turpin2018light}. 
Unfortunately, all existing networks are only trained and tested on the {\it same} diffuser, so the model may still be susceptible to speckle decorrelation.  
Indeed, as tested in our experiment, a single diffuser trained CNN does not capture sufficient statistical variations to interpret speckle patterns from other diffusers.  
Another closely related line of work is using DL to imaging through multi-mode fibers (MMF)~\cite{Borhani18,fan2018deep}.
Image transfer through a MMF also results in speckle patterns due to spatial mode mixing.  
CNNs have been designed to capture sufficient statistical variations of the setup so as to provide superior robustness against random variations.

We demonstrate our technique under shift-{\it variant} scattering by placing a diffuser at a defocused plane~\cite{mertz2015field,li2015conjugate,li2018imaging}.  
This geometry provides a limited isoplanatic region~($\approx$~speckle size)~\cite{mertz2015field,li2015conjugate}, as verified experimentally in Fig.~\ref{setup}.  
The objects extend well beyond the isoplanatic region ($\sim $300$\times$300 speckle size).   
Our task is further complicated by the intensity-only measurement under coherent illumination; the mapping between the object and speckle intensity is nonlinear~\cite{Goodman2007}. 
The training step in our DL method is conceptually similar to the TM calibration, in which a series of patterns are input to the diffuser and the output is measured.  
In TM calibration, interferometric measurements are often required~\cite{Popoff2010a,Kim2015}; additional phase-retrieval procedures are needed when intensity-only data are used~\cite{dremeau2015reference}.  
Here, the proposed CNN learns to interpret the `phaseless' measurements using its nonlinear, multilayer structure.

We experimentally achieve $\sim$256$\times$256-pixel SBP using up to 2400 training pairs.  
Importantly, our training data were collected on {\it multiple} diffusers.  
Distinct from the TM approach, our trained CNN  is able to predict objects through `unseen diffusers' that were {\it never used during training}.  
We experimentally quantify the CNN performance trained with 1, 2, or 4 diffusers and demonstrate the superior robustness over speckle decorrelation of our technique.    
We further demonstrate that the trained CNN is able to generalize over new object types through unseen diffusers.

Although it is hard to give an explicit expression of our CNN model (a common challenge in  DL), we attempt to provide some insights by performing both CNN visualization and statistical analysis on our data across multiple objects and diffusers.  
The basic mechanism of DL is to identify statistical invariance across large datasets~\cite{LeCun2015}.
We first visualize the activation maps of our CNN when inputting speckle patterns obtained from the same object but through different diffusers.  
By quantifying the correlations between the corresponding activation maps, we show that our CNN indeed gradually learns the invariance across these speckle patterns.
Next, we visualize speckle intensity correlations and show that physical invariance does exist across seemingly decorrelated speckle patterns taken through different diffusers. 
Such information would be hard to be directly utilized using existing models.  
Our CNN model is able to discover and exploit these `hidden' invariant features owing to its higher representation power.  
  
We demonstrate a promising DL framework towards  highly scalable imaging through scattering media.  Our method significantly improves the system's  information throughput and adaptability as compared to existing approaches, by  improving both the SBP and the robustness to speckle decorrelations.

\section{Method}
\label{section:method}

\subsection{Experimental setup}
\label{subsection:Experiment setup}

\begin{figure}[t]
\centering\includegraphics[width=1\linewidth]{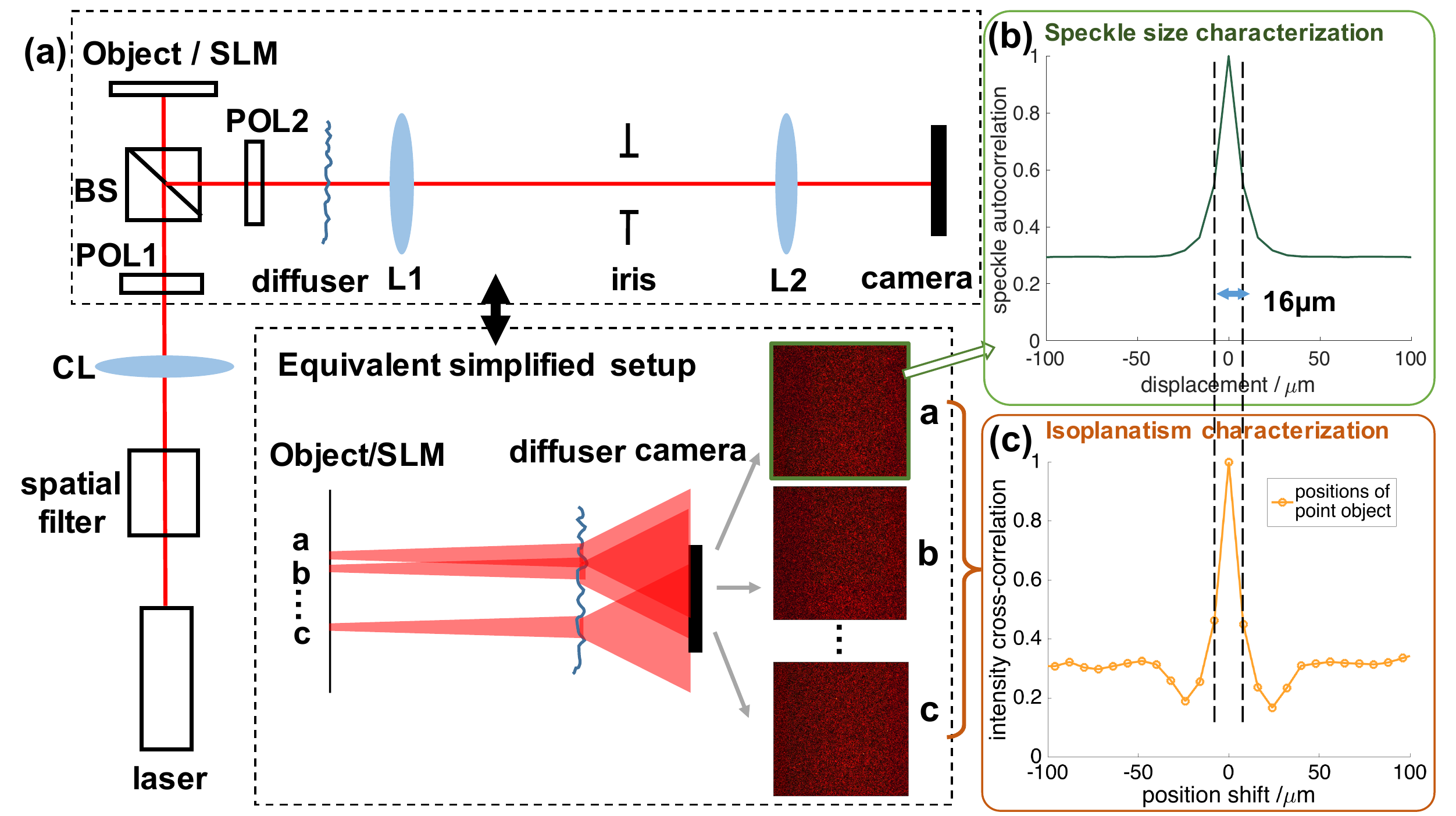}
\caption{(a) Experiment setup uses an SLM as the object that is  illuminated by a laser.  A diffuser is placed at a defocused plane to create shift-{\it variant} scattering. (b) The speckle size is $\sim$16$\mu m$, characterized by the speckle's intensity autocorrelation. (c) The isoplanatic range is $\sim$1 speckle size  characterized by the cross-correlation coefficients between speckle patterns of shifted point objects.
}
\label{setup}
\end{figure}

We use a spatial light modulator (SLM) (Holoeye {NIR-011}, pixel size 8$\mu$m) as a programmable {\it amplitude}-only object with two orthogonally oriented polarizers before and after [Fig.~\ref{setup}(a)], similar to~\cite{li2018imaging}. 
It is coherently illuminated by a collimated beam from a HeNe laser (632nm, Thorlabs HNL210L). 
The SLM is relayed onto the camera (Thorlabs Quantalux, pixel size 5.04$\mu$m) by a 4F system.   
Two lenses with focal lengths 200mm (L1), and 125mm (L2) are used to provide a 0.625 magnification.  
This design approximately produces the same effective pixel size for the object and the image, which is convenient for the CNN implementation since the same number of pixels can be used for the input and output without resizing~\cite{li2018imaging}.  
Precise pixel-wise alignment was {\it not} performed nor needed.  
A $\sim$9mm iris is placed at the pupil plane of the 4F system to control the speckle size. 
The theoretical average speckle size is $\sim$8.8$\mu$m, or equivalently $\sim$14$\mu$m on the object plane, as set by $\lambda /${2NA} (NA denotes the numerical aperture of the 4F system)~\cite{Goodman2007}. 
This is experimentally verified by taking the autocorrelation of a speckle pattern through a  diffuser and measuring the full-width at half-maximum~\cite{Goodman2007}, which reads $\sim$16$\mu$m, as shown in Fig.~\ref{setup}(b) (for ease of comparison, all length measurements are converted to the object side). 

\begin{figure*}[h]
\centering\includegraphics[width=\textwidth]{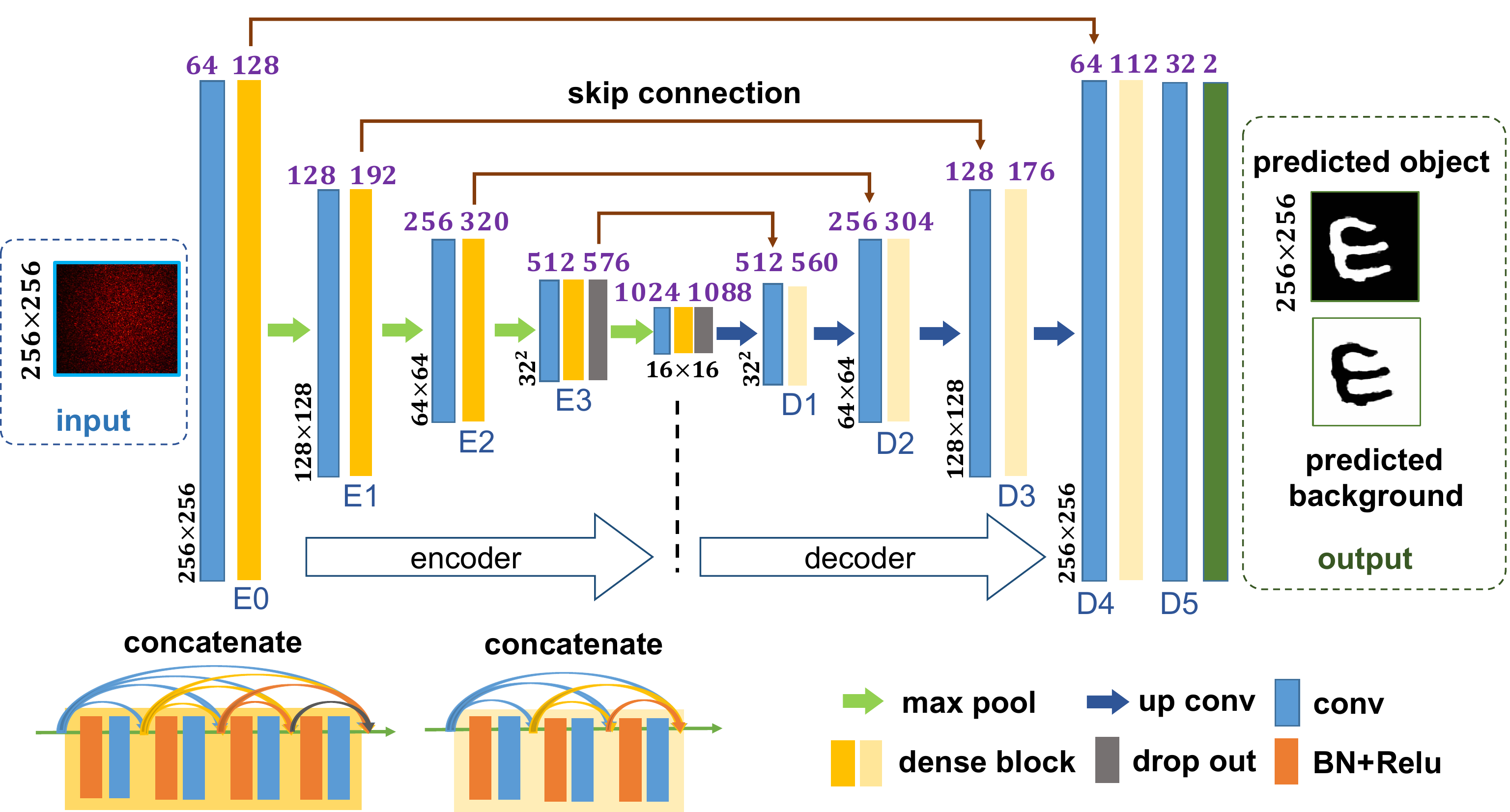}
\caption{The proposed CNN architecture to learn statistical relationship between speckle patterns and unscattered object. It takes the general encoder-decoder Unet structure (the  layer indices are marked in blue).  Starting with a high-resolution input speckle pattern, the encoder gradually condenses the lateral spatial information (size marked in black) into high-level feature maps with growing depths (size marked in purple); the decoder reverses the process by recombining the information into feature maps with gradually increased lateral details; the output consists of a two-channel object, background pixel-wise prediction.
}
\label{fig:structure}
\end{figure*}

The spatially {\it variant} scattering is generated by placing a thin glass diffuser (Thorlabs, 220 grits, DG10-220) between the SLM and the 4F system's first lens.  
This system theoretically provides a small isoplanatic region that is limited to a single speckle since the diffuser is placed at a defocus position~\cite{mertz2015field}. 
We quantify the isoplanatism by measuring the intensity speckle correlations~\cite{Katz2014}.  
A 3$\times$3 pixel `point-object' is scanned linearly across the SLM pixel-by-pixel ($8\mu$m).  
The isoplanatic range is then found by calculating the PCC between the speckle pattern from the central point and the one from each shifted point. 
Rapid speckle decorrelation beyond a single speckle range is observed in Fig.~\ref{setup}(c).  
The correlation coefficient plateaus around 0.3, close to the value in the speckle intensity autocorrelation curve [Fig.~\ref{setup}(b)].  
The smallest object ($24\mu$m) was limited by the signal-to-background ratio of the experiment due to imperfect polarizer extinction power producing non-negligible background at low-light levels. 
The same procedure was repeated on different object sizes; nearly identical curves are obtained (see the supplementary material).
The same behavior was numerically predicted in~\cite{mertz2015field,li2018imaging}.

Speckle measurements are repeated on several diffusers having the same macroscopic parameter (220 grits).  
All glass~\mbox{(BK-7)} diffusers are manufactured by the same process (Thorlabs), in which the top surface is first polished, and the bottom surface is then ground with the specified (220) grit.
220-grit provides an average 63$\mu$m feature size  on the glass surface. 
When imaged in our setup, speckles generated by all diffusers possess similar statistical properties, including the average speckle size, and the background correlation (0.3) (see Figs.~\ref{setup} and \ref{fig:correlation}).

\subsection{Data acquisition}
\label{subsection:acquisition}

The central 512$\times$512 SLM pixels  are used as the object; the corresponding central 512$\times$512 camera pixels are used as the speckle intensity for CNN training and testing. 
Considering the system's resolution (measured by the speckle size), the SBP is $\sim$300$\times$300 pixels with a field-of-view (FOV) of $\sim$4$\times$4 mm$^2$, which is well beyond the isoplanatic patch.  
The objects displayed on the SLM are 8-bit grayscale images from the MNITS handwritten digit~\cite{url:MNIST}, NIST handwritten letter~\cite{url:MIST}, and Quickdraw objects~\cite{url:quickdraw} databases. 

In total we take speckle patterns using 9 different diffusers.
We use data from up to 4 diffusers to train our CNN, {\it the data from the other 5 diffusers are never seen by the CNN during training and are only used for testing}.  
The {\it training} objects are only taken from the handwritten digit and letter databases.  
The Quickdraw objects are only used for {\it testing}. 
The data were taken spanning $\sim$8 weeks, demonstrating the robustness of our approach to possible random variations during the experiment.

To collect the {\it training} data, we use in-total 600 objects (300 digits and 300  letters).  For each {\it training} diffuser, we take 600 speckle images, giving in-total up to 2400 training dataset.

Our {\it testing} data are purposely designed to have four groups for characterizing our CNN's generalization capability evaluated from different perspectives:

\noindent{\bf Group 1}  tests the CNN generalization over {\it speckle decorrelation due to the change of diffusers.}
It consists of 3000 {\it `seen objects through unseen diffusers'} collected from the {\it same} 600 objects used in the training, but through the 5 {\it unseen testing} diffusers.  

\noindent{\bf  Group 2} tests the CNN over {\it the change of diffusers and unseen objects of the same type} (as the training objects).
It consists of 200 {\it `unseen objects of the same type through unseen diffusers'} from previously {\it unused} 200 objects during the training and of the same class (100 digits + 100 letters), and through a randomly selected {\it unseen testing} diffuser.  

\noindent{\bf  Group 3} tests the CNN over {\it the change of diffusers and new object types.}
It consists of 800 {\it `unseen objects of new types through unseen diffusers'}, and through the 5 {\it unseen testing} diffusers.  
The objects are taken from the Quickdraw database.

\noindent{\bf  Group 4}  benchmarks the CNN performance {\it trained on a single diffuser}.  
 It consists of 28 {\it `unseen objects through the same diffuser'} from previously {\it unused} 28 objects of the same type (9 digits + 19 letters) during training, and through a randomly selected {\it seen training} diffuser.

\subsection{Data preprocessing}
 
Due to computational limitations, all input and output images are first downsampled from 512$\times$512 pixels to 256$\times$256 pixels by taking the average within each 2$\times$2 neighboring pixels (i.e. 2$\times$2 binning).  
The downsampling reduces both the number of network parameters (which grows with the input size) and the required data size for training without overfitting (which grows with the network parameters). 
However, two artifacts may be resulted. 
First, our system images each speckle with approximately two pixels; after downsampling, each binned {\it image} pixel contains intensities from several speckle grains, effectively reducing the  contrast of the input patterns~\cite{Goodman2007}. 
Second, each binned {\it object} pixel may combine pixels from both the object and background regions, introducing incorrect (noisy) ground-truth.  
Robust training using noisy ground-truth has been shown in other CNN tasks~\cite{xiao2015learning}. 
In essence, the CNN learns the invariants and filters out the random noise. 
Our results suggest that the downsampling has little effect to the final results.
Next, for both training and testing, the input speckles  are normalized between 0 and 1 by dividing each image by its maximum.  

Our CNN is designed to perform two types of tasks. 
First, the  {\it binary detection} task outputs a two-channel binary estimate of the object and background.  
Accordingly, during the training, each grayscale object is thresholded by setting all non-zero valued pixels to 1 to give the ground-truth object; the ground-truth background is the complement. 
Second, the {\it grayscale object reconstruction} task outputs a two-channel grayscale estimate of the object and background. 
The ground-truth object is the grayscale image displayed on the SLM, processed with 2$\times$2 pixel binning and normalized between 0 and 1; the ground-truth background is defined by subtracting the ground-truth object from 1.

\subsection{CNN implementation}

We build a CNN to learn a statistical model relating the speckle patterns and the unscattered objects. Importantly, the goal is to make predictions through previously unseen diffusers. 

The overall structure of the proposed CNN (Fig.~\ref{fig:structure}) follows the encoder-decoder `Unet' architecture~\cite{ronneberger2015u} with modifications of replacing each  convolutional layer with a dense block~\cite{huang2017densely} to improve the training efficiency~\cite{li2018imaging}. 
The input to the CNN is a preprocessed 256$\times$256 speckle pattern. 
Next, the input goes through the `encoder' path, which consists of 4 dense blocks connected by max pooling layer for downsampling.  
The intermediate output from the encoder has small lateral dimensions (16$\times$16), but encodes rich information along the `depth' (having 1088 activation maps).  
Each dense block contains multiple layers, in which each layer consists of batch-normalization (BN), the rectified linear unit (ReLU) nonlinear activation, and convolution (conv) with 16 filters. 
Next, the low-resolution activation maps go through the `decoder' path, which consists of 4 additional dense blocks connected by up-sampling convolutional (up conv) layers. 
The information across different spatial scales are tunneled through the encoder-decoder paths by skip connections to preserve high-frequency information. 
After the decoder path, an additional convolutional layer followed by the last layer produces the network output.  
The design of this last layer requires careful consideration of {\it the desired imaging task}.   

Our CNN is designed to {\it image sparse objects}.  
Widely used loss functions including mean squared error (MSE) and mean absolute error (MAE), cannot promote sparsity since they assume the underlying signals follow Gaussian and Laplace statistics, respectively~\cite{kendall2017uncertainties}.
In a recent work~\cite{li2018imaging}, the negative PCC is shown to promote sparse  predictions.  
Here, we propose an alternative method.  
First, we use a softmax layer to produce a pair of mutually complementary object  and  background channels.  
We then use the averaged cross-entropy~\cite{ronneberger2015u} as the loss function~$L$,  which has shown to promote sparsity~\cite{suresh2008risk}, and is given by 
\begin{equation}
L = \frac{1}{2N}\sum_c\sum_x -(g\log(p)+(1-g)\log(1-p))
\label{eq:lossfunction1}
\end{equation}
where $g$ is the ground-truth pixel value and $p$ represents the prediction; the average is over all $N$-pixels~$x$ across both channels~$c$.  
Both $g$ and $p$ can take binary or continuous values.

Importantly, our design allows making both binary and grayscale predictions.  
First, we consider {\it the pixel-wise binary detection} problem -- the CNN predicts if the object is present or not pixel-by-pixel.   
In this case, both the ground-truth and predictions take binary values.  
The intermediate output from the softmax layer is often interpreted as the probabilities of each pixel belonging to the object and background classes.
Second, we consider {\it the grayscale object reconstruction} problem -- the CNN predicts  continuous-valued intensity in each object pixel. 
In this case, both the ground-truth and predictions take grayscale values.  
The predictions are directly from the softmax layer. 
Since our objects are generated with a 8-bit SLM, the CNN predictions are set to the same bit-level.

The CNN training was performed on the BU SCC with one GPU (NVIDIA Tesla P100) using  Keras/Tensorflow. 
Each CNN is trained with 500 epochs by the Adam optimizer for up to 44 hours.  
The learning rate of $10^{-4}$ is used for the first 300 epochs, $10^{-5}$ for the next 100 epochs, and $10^{-6}$ for the final 100 epochs.  
Once the CNN is trained, each prediction was made in real-time. 
More details of the CNN architecture, parameter optimization, and training procedures are provided in the supplementary material.  
We also provide open source code of our CNN model along with pre-trained weights and sample data in~\cite{url:deepspeckle}.

\section{Results}
\label{sec:result}

\begin{figure}[h]
\centering\includegraphics[width=1\linewidth]{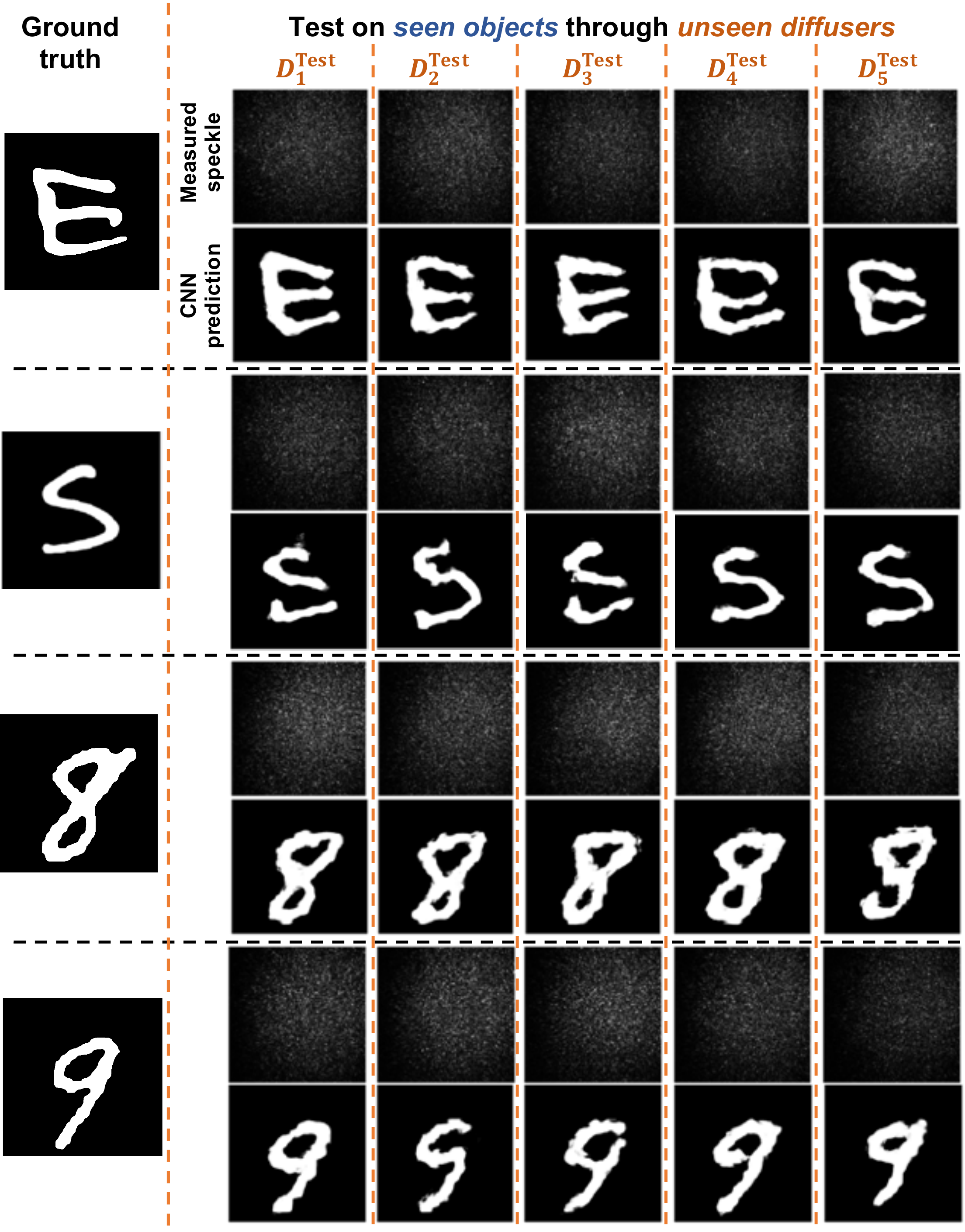}
\caption{Testing results of `seen objects through unseen diffusers'.
The {\it single}  CNN trained with four `training diffusers' is used to predict objects through previously unseen `testing diffusers', $D^{\mathrm{test}}_1,D^{\mathrm{test}}_2, D^{\mathrm{test}}_3,D^{\mathrm{test}}_4,D^{\mathrm{test}}_5$.  The same set of objects are used during the training through the training diffusers.
Despite the apparent differences across the speckle patterns, consistently reliable predictions are made by our CNN.}
\label{fig:result}
\end{figure}

We present our results from four types of experiments, in line with the acquired data described in Sec.~2\ref{subsection:acquisition}.  
The results from the first three experiments are all from the CNN trained with 4 training diffusers and tested on 5 testing diffusers. 
The last experiment is to compare the 4-training-diffuser results against those  from the CNN trained on a single diffuser.  
Although our CNN is able to make both binary and grayscale predictions, we here only show binary images.  
Grayscale network provides similar performance, as detailed in the supplementary material.
This is probably because our CNN is designed to image sparse objects.  
Imaging non-sparse objects become more challenging~\cite{li2018imaging}, which will be considered in our future work.

\begin{figure}[!t]
\centering\includegraphics[width=0.8\textwidth]{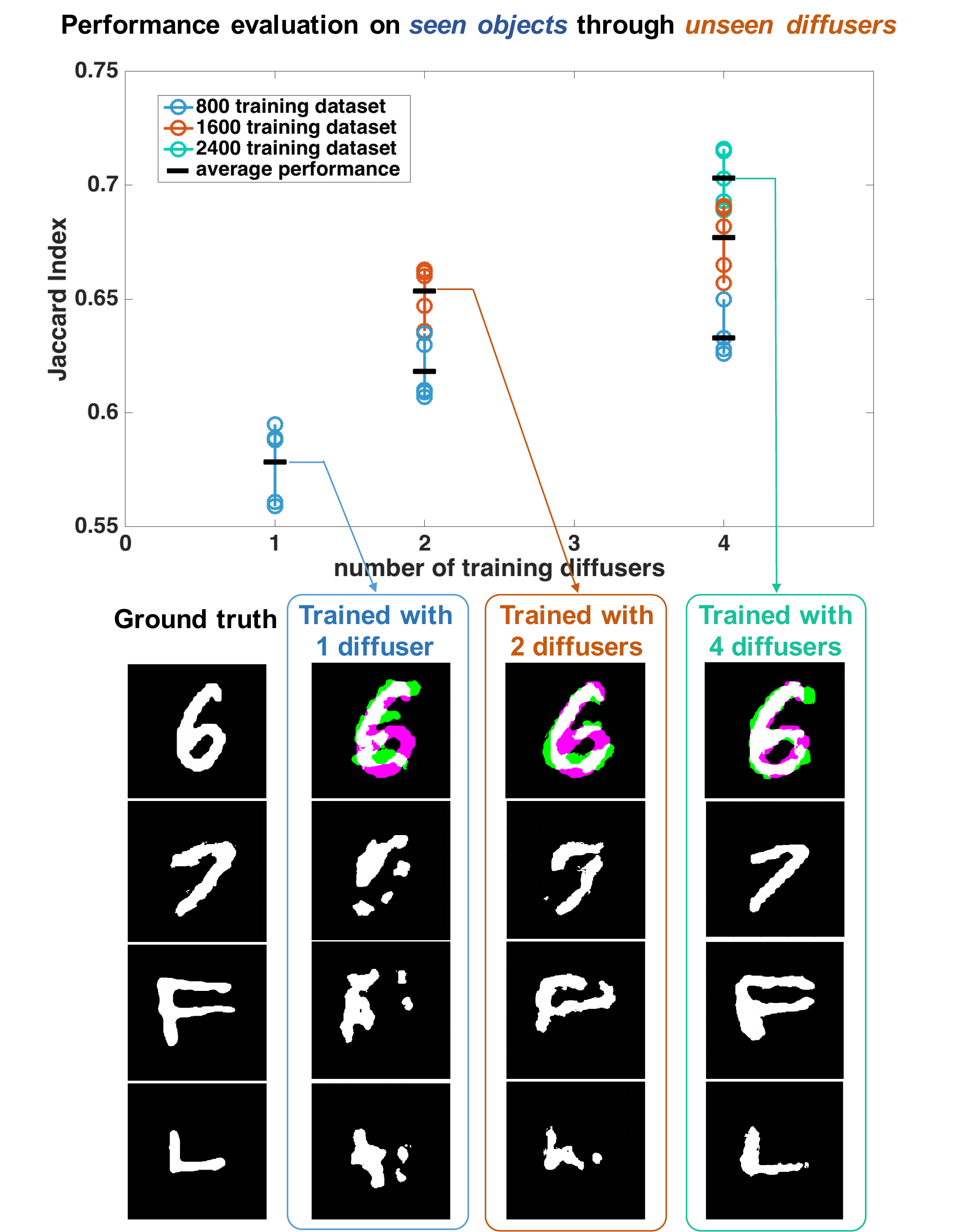}
\caption{We compare the performance of multiple CNNs trained on 1, 2, and 4 diffusers using different dataset sizes (800 in blue, 1600 in orange, 2400 in green) by the Jaccard index (JI). 
Each CNN is tested under the same condition, using the same 1000 speckle patterns from seen objects through 5 unseen diffusers. 
Each circle represents the average JI on all objects through each testing diffuser. 
The mean JI of each CNN is marked by black horizontal bars.
The bottom figure shows representative example predictions from the CNN trained on 1, 2, and 4 diffusers, respectively.
To visualize the result, the first row shows the CNN prediction that is overlaid with the true-positive (white), the false-positive (green), and the false-negative (purple). }
\label{fig:metric}
\end{figure}

In the first experiment, we test our CNN to {\it predict `seen objects through unseen diffusers'} ({\bf Task 1}). 
Notably, our CNN demonstrates superior generalization in predicting objects through previously unseen diffusers. 
Representative examples of the speckle and prediction pairs are shown in Fig.~\ref{fig:result}.  
More results are given in the supplementary material.
For the same object, although the speckle patterns through different diffusers appear notably different, the CNN consistently makes high-quality predictions.  
Later, we quantify the differences between these speckle patterns  by speckle decorrelation analysis in Sec.~\ref{sec:analysis}.
The prediction results present slight variations since our CNN  makes {\it pixel-wise predictions}, rather than the whole-image classification~\cite{Krizhevsky.etal2012}. 
Our pixel-wise prediction task is considerably more difficult since the network needs to effectively learn the per-pixel input-output relation.
In addition, since our CNN adapts to {\it all diffusers of the same class}, the learned relation needs also to be adaptable to all possible statistical variations.
The variations of the predictions for this task using our binary CNN are quantified later in Fig.~\ref{fig:metric}.
Representative examples and statistical analysis on the grayscale CNN predictions are provided in the supplementary material.

\begin{figure}[h]
\centering\includegraphics[width=0.96\linewidth]{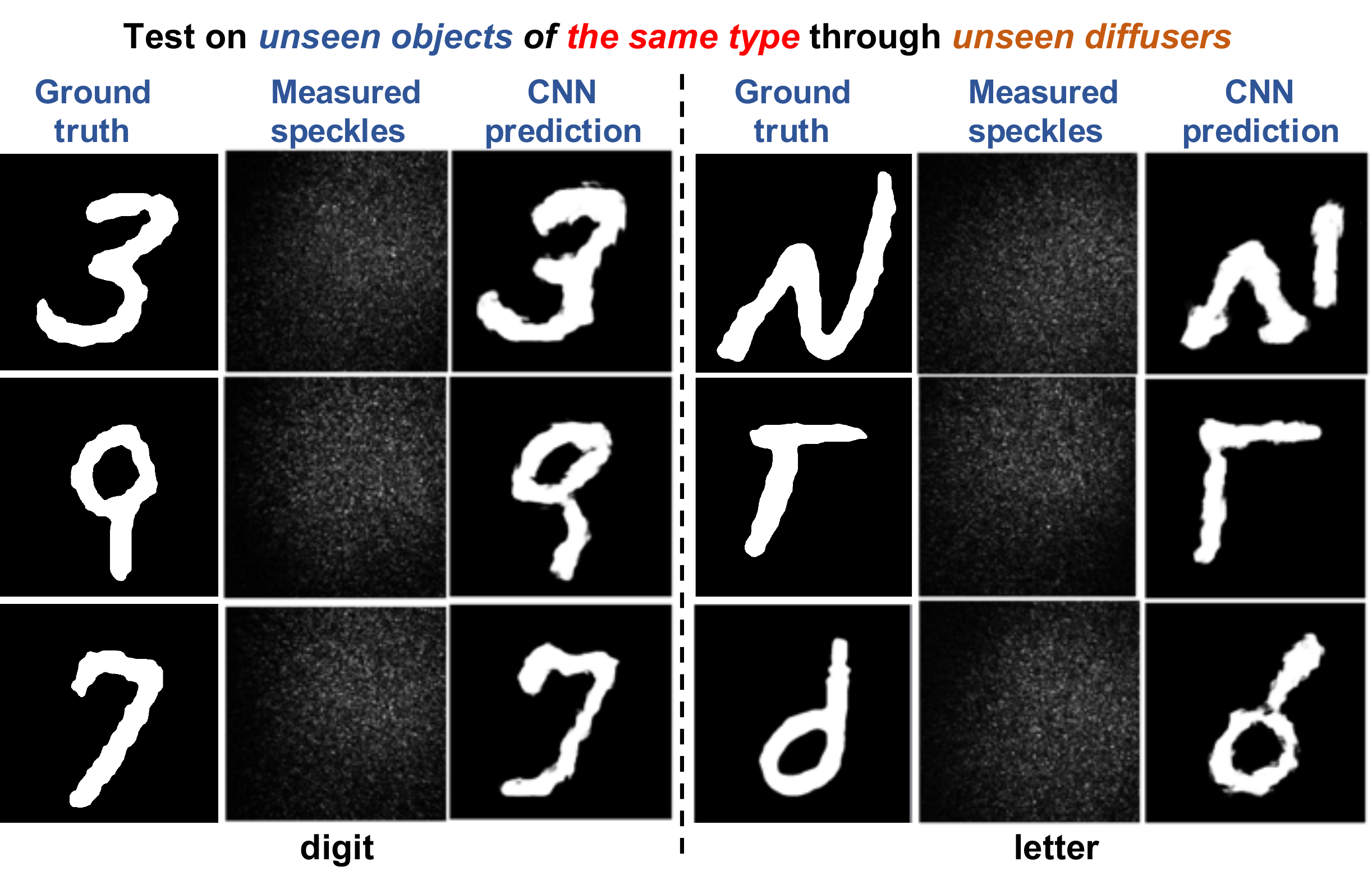}
\caption{Testing results of `unseen objects of the same type through unseen diffusers'. 
The CNN trained with four `training diffusers' is used to make predictions using speckles from previously unused objects (during training) through unseen testing diffusers.
The testing objects belong to the same class (handwritten digits and letters) as the training sets.}
\label{fig:result2}
\end{figure}

In the second experiment, we test our CNN on a more difficult task of {\it predicting `unseen objects of the same type through unseen diffusers'}~({\bf Task 2}).  
The set of objects have never been used in the training.  
They, however, belong to the same object class to the training data, i.e. handwritten digits and letters. 
A quantitative comparison between Task 1 and Task 2 measured by the speckle decorrelation is presented in Sec.~\ref{sec:analysis}.
Representative examples are shown in Fig.~\ref{fig:result2}, demonstrating that the CNN is able to make high-quality binary predictions of these {\it unseen objects from the same class}, while through unseen diffusers. 
The corresponding grayscale predictions are shown in the supplementary material.

\begin{figure}[t]
\centering\includegraphics[width=1\linewidth]{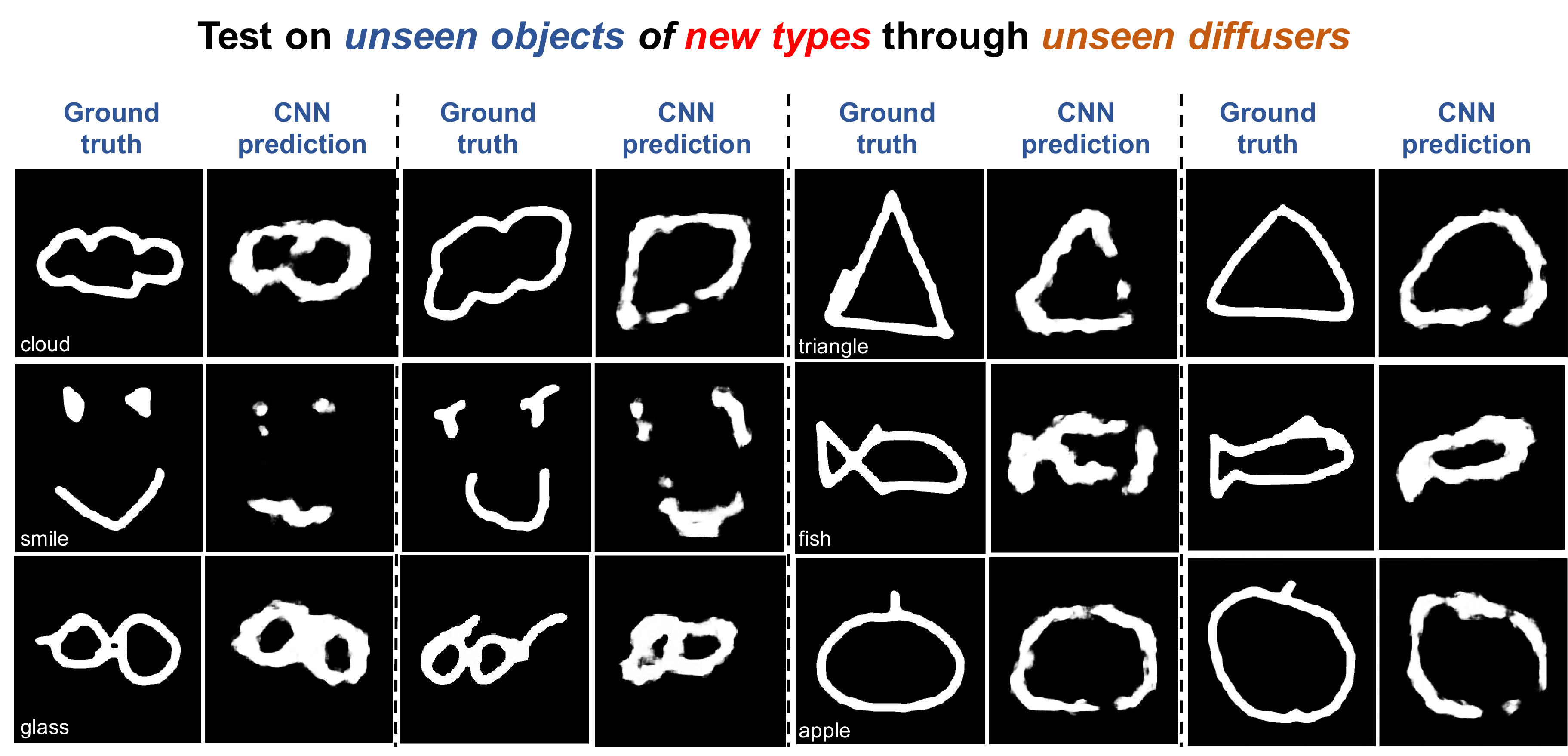}
\caption{Testing results of `unseen objects of new types through unseen diffusers'. 
The CNN trained with four `training diffusers' is used to make predictions using speckles from new types of objects through unseen testing diffusers.
The testing objects are taken from a new class (Quickdraw) that have never been used during training.
}
\label{fig:result_new}
\end{figure}

\begin{figure}[t]
\centering\includegraphics[width=0.6\textwidth]{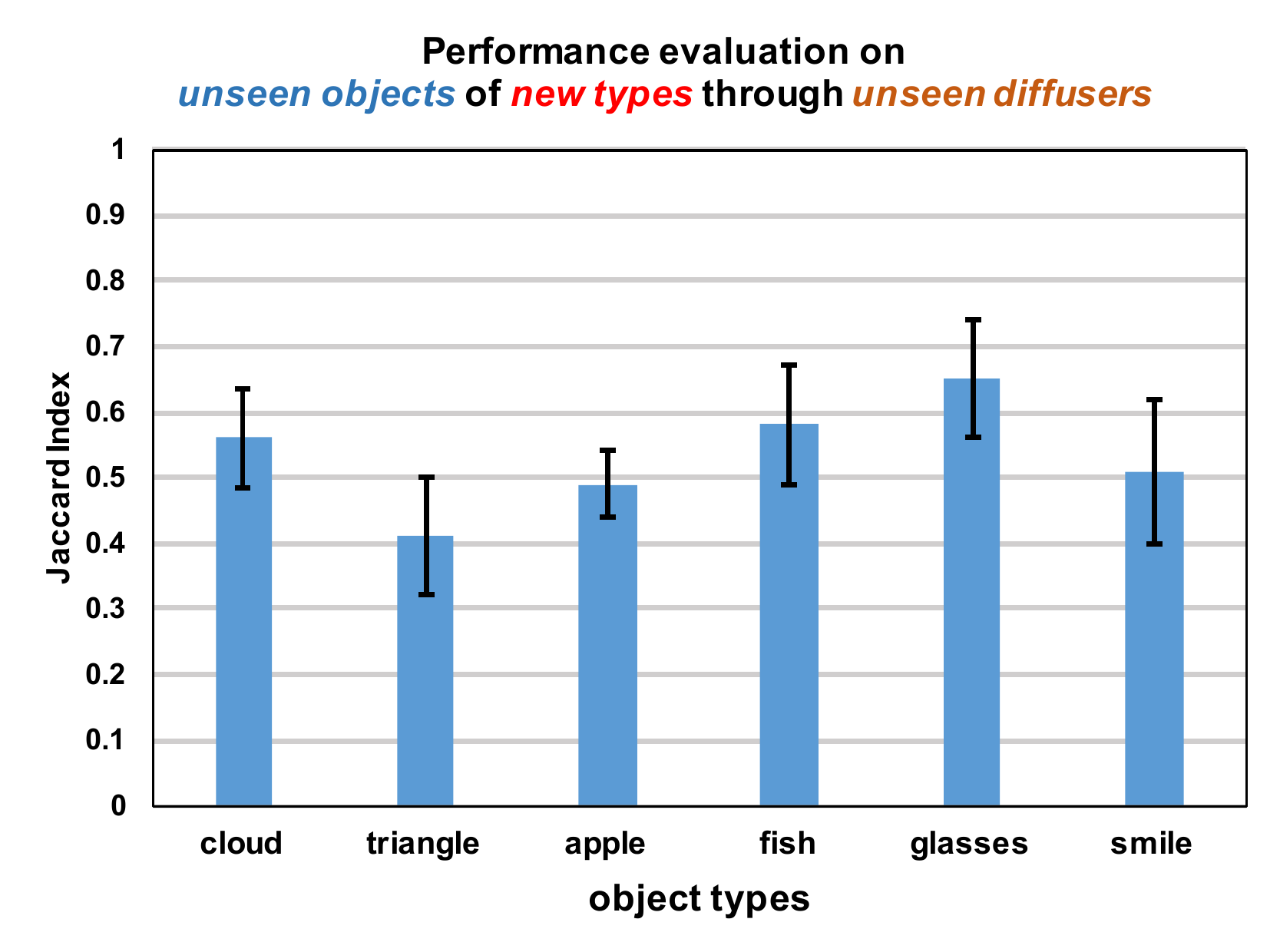}
\caption{Quantitative evaluation of the CNN performance on `unseen objects of new types through unseen diffusers'.
Each bar represents the mean Jaccard index (JI) from different objects  belonging to the same type and are imaged through 5 different testing diffusers. 
Each error bar represents the standard deviation of the JI for each object type. }
\label{fig:metric_new}
\end{figure}

In the third experiment, we further test our CNN on {\it predicting `unseen objects of new types through unseen diffusers'}~({\bf Task 3}).  
The set of objects have never been used in the training and belong to a {\it different object class} (Quickdraw). 
Representative examples are shown in Fig.~\ref{fig:result_new}, demonstrating that our  CNN is still able to make high-quality predictions of these {\it unseen new types of objects} through unseen diffusers. 
The quality of the binary predictions for this task are quantified in Fig.~\ref{fig:metric_new} across different object types.
The corresponding grayscale predictions are evaluated in the supplementary material.

\begin{figure}[h]
\centering\includegraphics[width=0.96\linewidth]{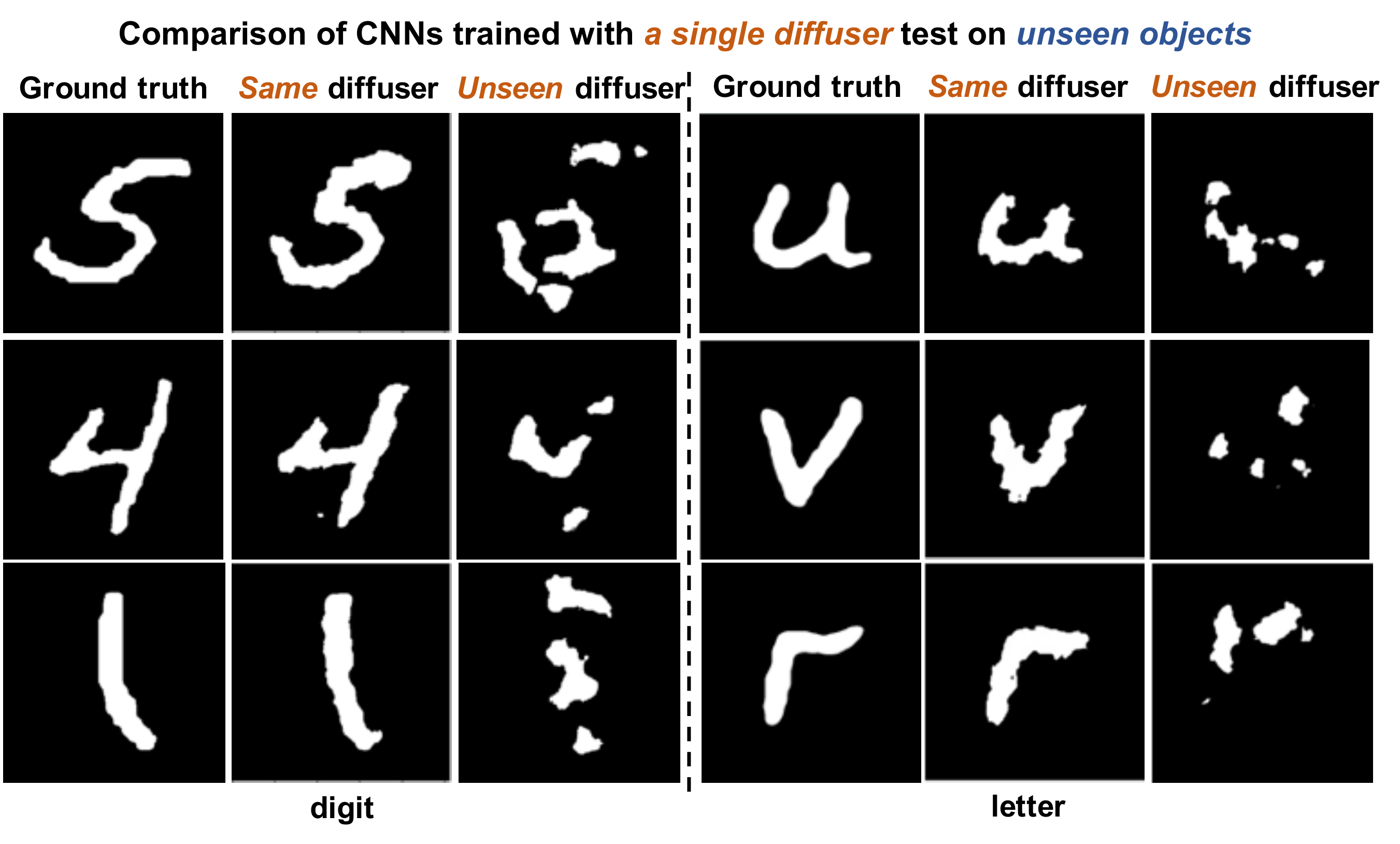}
\caption{Testing results of the CNN trained on {\it a single diffuser}.  When tested on speckles from unseen object (during training) through the {\it same} diffuser, the CNN is able to make high-quality predictions.  
However, it fails on speckles from a different unseen diffuser, demonstrating the importance of the proposed DL strategy involving multiple diffusers.}
\label{fig:result3}
\end{figure}

In the fourth experiment, we compare our `4-training-diffuser' results against those from {\it the CNN trained on a single diffuser}.  
The results are presented in Fig.~\ref{fig:result3}, which consists of two tasks. 
{\bf Task 4} makes predictions on {\it unseen objects by the CNN that is trained and tested on the same diffuser}.
Successful demonstrations of accomplishing this task via machine learning have been reported~\cite{horisaki2016learning,lyu2017exploit,li2018imaging}. 
{\bf Task 5} makes predictions on {\it unseen objects through a different unseen diffuser by the CNN trained on a single diffuser}.
The goals of this experiment are in twofolds.  
First, due to the different choices of CNN architectures and loss functions, here we validate that our design can indeed reliably perform Task 4, as shown in Fig.~\ref{fig:result3}.  
Our results from our CNN are further quantified in Fig.~\ref{fig:metric}, which match the state-of-the-art performance with an average PCC of 0.626~\cite{li2018imaging}.   
Second, we verify that a CNN trained on only a single diffuser can{\it not} be reliably generalized to other diffusers (shown in Fig.~\ref{fig:result3}), since the CNN is tuned to only fit to the model of a specific diffuser.

Next, we quantify the performance on the `seen objects through unseen diffusers' task.
We expand the comparisons across 6 CNNs trained on 1,2, or 4 diffusers with 3 training dataset sizes (in total: 800, 1600, and 2400 pairs).  
We use two metrics, including the Jaccard index (JI) and PCC.  
Both metrics are useful to measure the similarity between image pairs~\cite{zou2004statistical}; they provide slightly different scores due to the differences in error-counting. 
Each CNN is tested under the same condition, using the same 1000 speckle patterns (the  same as Fig.~\ref{fig:result}).

We first present the JI scores. 
In the top figure of Fig.~\ref{fig:metric}, the JI of each CNN tested on each individual testing diffuser is shown as a circle.  
Results from all 5 unseen diffusers are clustered together, regardless of the CNN being used, demonstrating {\it the consistency of the CNN prediction against object and diffuser variations}.  
In addition, we make two observations.
First, {\it the performance improves as more training diffusers are used.} 
This is evident by comparing the results from the same number of 800 training dataset while increasing the number of training diffusers (similarly for the 1600 case).
Second, {\it the performance further improves by increasing the size of training dataset.}
This is seen by comparing  the same number of 4 training diffusers while increases the training dataset size (similarly for the 2-diffuser case). 
To provide an intuitive visualization of the JI score, the bottom figure of Fig.~\ref{fig:metric} shows a few representative examples. 
In the first row, the result is further broken down to the true-positive (white), the false-positive (green), and the false-negative~(purple). 

\begin{table}[!h]
  \begin{center}
    \caption{Pearson correlation coefficient of CNN predictions for the `seen objects through unseen diffusers' task.}
    \label{table1}
    \begin{tabular}{l|c|c|r} 
      ~ & \textbf{1 diffuser} & \textbf{2 diffusers} & \textbf{4 diffusers} \\
      \hline
      800 dataset & 0.429 & 0.473 & 0.568  \\
      1600 dataset&  & 0.528 & 0.577\\
      2400  dataset &  &  & 0.626\\
    \end{tabular}
  \end{center}
\end{table}

Next, we provide the alternative evaluation using the PCC score. 
The mean PCC of each CNN is given in Table~\ref{table1}. 
The general observations remain the same as the JI evaluation. 
In addition, we observe that the performance from `4 diffusers, 800 dataset' is slightly better than that from '2 diffusers, 1600 dataset' (i.e. more diffusers and less dataset), further demonstrating the effectiveness of training using multiple diffusers.  

Finally, we quantify the performance on the `unseen objects of new types through unseen diffusers' task in Fig.~\ref{fig:metric_new}.  
The results are from the CNN trained with 4 training diffusers and 2400 training datasets (the condition for Fig.~\ref{fig:result_new}).  
In general, our trained CNN is able to make high-quality predictions albeit with reduced JI scores as compared to the `seen object' case. 
The performance also varies with the specific object types.
In total, we tested 6 different types, whose performance are quantified by the mean and standard deviations of the JI.
These results suggest that the quality of the CNN model is also influenced by the object types used during training.
A larger training dataset covering additional object types may further improve our results.

\section{analysis}
\label{sec:analysis}

To provide some insights of our CNN model, we perform analysis on both the network and the speckle patterns.
The main principle of DL is to learn statistical invariant information across large dataset~\cite{LeCun2015}.
Thus, our goal is to look for {\it any meaningful invariant features} among speckles taken through different diffusers.  
If found any, it can suggest that it is plausible to establish a statistical mapping to relate these speckles by the CNN model.

\begin{figure*}[t]
\centering\includegraphics[width=1\linewidth]{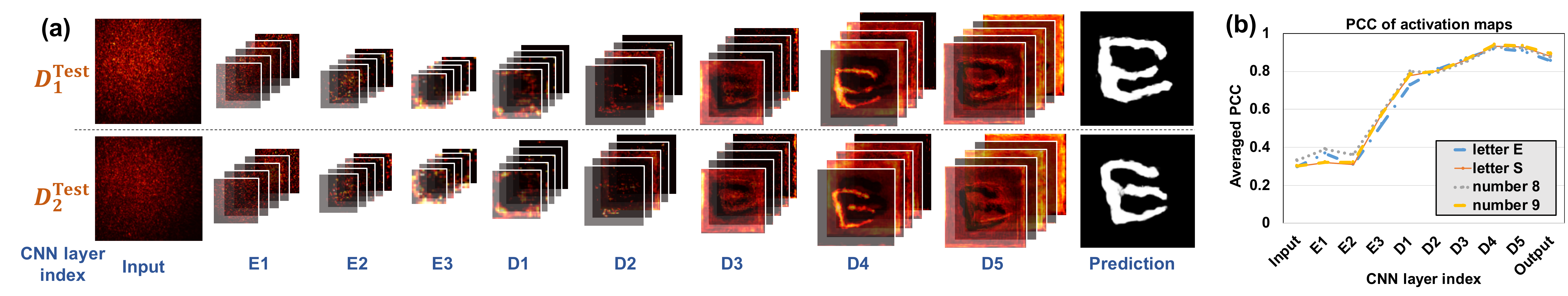}
\caption{(a) The visualization of the intermediate activation maps (layer index defined in Fig.~\ref{fig:structure})
of our trained CNN by inputting two speckle patterns from {\it the same object}, through {\it two different testing diffusers}. 
The number of channels in each layers are defined by the depths of each layer in Fig.~\ref{fig:structure}.
The corresponding activation maps from the two speckle patterns become increasingly similar as the data flow through deeper layers, demonstrating the CNN's ability to extract statistically invariant information from visually distinct speckle patterns.
(b) The similarity of the activation maps are quantified using the Pearson correlation coefficient (PCC) averaged over all possible pairs of data from the 5 testing diffusers using the same object. 
The PCC generally grows with the layer index.
Results from four different objects are shown, all of which follow the similar trend.
}
\label{fig:activation}
\end{figure*}

First, we visualize the intermediate activation maps~\cite{zeiler2014visualizing} from each layer of our CNN when inputting speckle patterns from {\it the same object} but through {\it different testing diffusers}. 
Starting with a pair of visually distinct speckle patterns, the activation maps gradually resemble similar patterns as the data flow through the encoder-decoder paths, as shown in Fig.~\ref{fig:activation}(a).  
To quantify the learned invariance, we compute the pair-wise PCCs of each corresponding layer (across all channels) from the same object for all possible combinations of the 5 testing diffusers.  
The PCC generally grows as the CNN layer; PCC curves from different objects follow  the similar trend, as shown in Fig.~\ref{fig:activation}(b).

\begin{figure*}[h]
\centering\includegraphics[width=\textwidth]{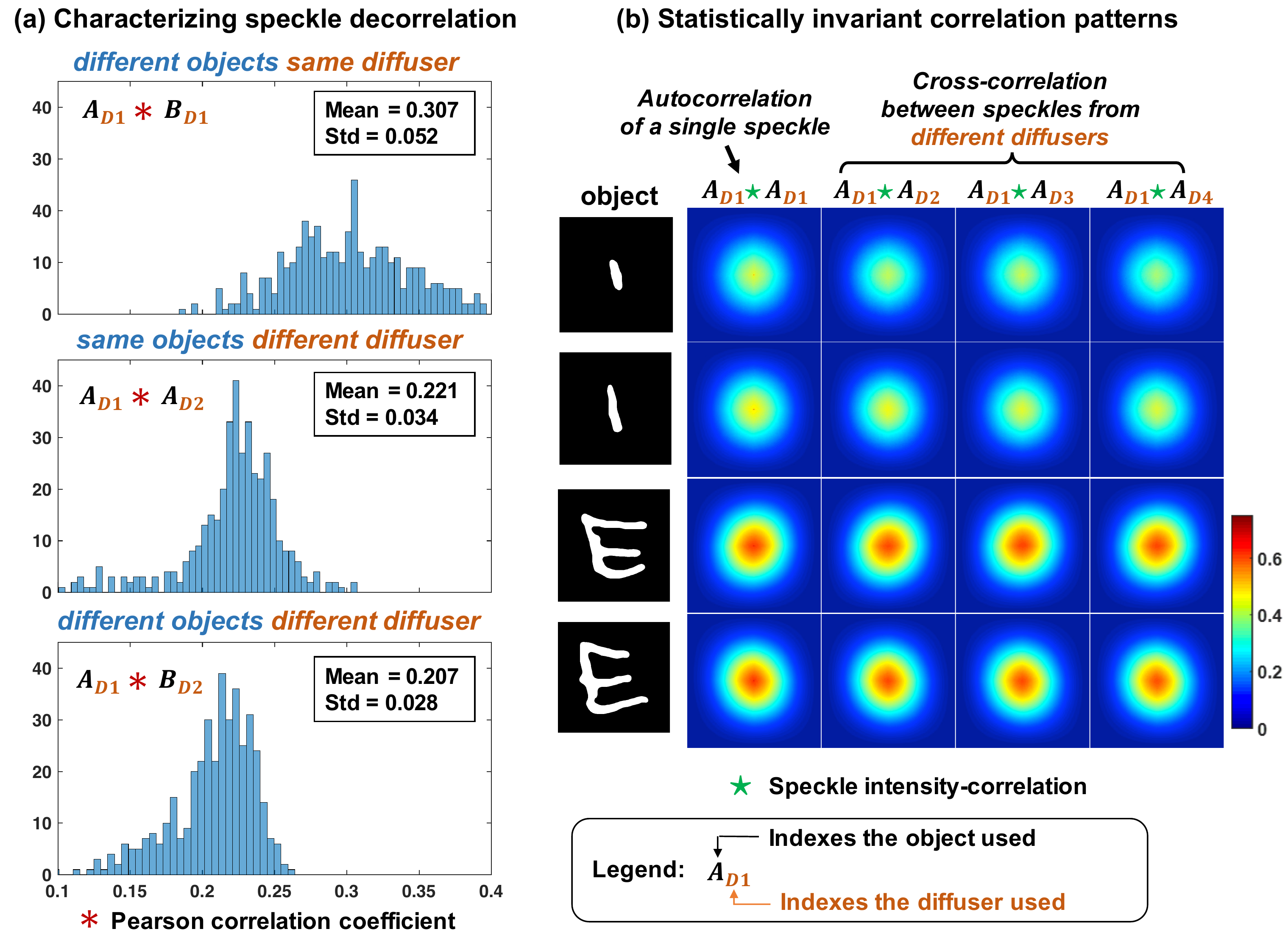}
\caption{(a) To quantitatively analyze the robustness of our CNN to speckle decorrelations, Pearson correlation coefficients are calculated based on randomly selected 400 speckle patterns for three different cases: $A_{\mathrm{D1}}*B_{\mathrm{D1}}$: different objects, the same diffuser; 
$A_{\mathrm{D1}}*A_{\mathrm{D2}}$: the same object, different diffusers; 
$A_{\mathrm{D1}}*B_{\mathrm{D2}}$: different objects, different diffusers.  
The results show progressively more difficult tasks tested in our experiments. 
Training and testing on the same diffuser (Fig.~\ref{fig:result3}) needs to overcome an average 0.307 decorrelation; training on one diffuser and testing on another diffuser but with the same object (Fig.~\ref{fig:result}) needs to account for an average 0.221 decorrelation; 
training and testing on different objects and diffusers (Figs.~\ref{fig:result2}, \ref{fig:result_new}, and \ref{fig:result3}) needs to further model an average 0.207 decorrelation. 
(b) Correlating speckle patterns from the same objects but through different diffusers shows invariant patterns, which provides a possible source of weak correlation information exploited by the CNN.}
\label{fig:correlation}
\end{figure*}

Next, we perform speckle correlation analysis. 
Our findings are summarized in Fig.~\ref{fig:correlation}.  
First, we quantify speckle decorrelation in our measurement using the classical PCC metric~\cite{hillman2013digital,jang2015relation,liu2015optical}.  
Figure~\ref{fig:correlation}(a) presents the PCC's histograms under various tasks (defined in Sec.~\ref{sec:result}), each from 400 randomly chosen speckle patterns. 
We describe the result based on the order of decorrelation (hence the difficulty of the task). 
First, Task 4 (Fig.~\ref{fig:result3}) is evaluated by $A_{\mathrm{D1}}*B_{\mathrm{D1}}$, which correlates speckles from {\it different objects through the same diffuser}. 
Most of the speckle patterns are decorrelated and the mean coefficient is 0.307,  which is consistent with the values found in both the isoplanatism  and speckle size characterization plots in Fig.~\ref{setup}. 
Second, Task 1  (Fig.~\ref{fig:result}) is evaluated by $A_{\mathrm{D1}}*A_{\mathrm{D2}}$, which is for {\it the same object through different diffusers}.
The speckle patterns are further decorrelated to a mean value of 0.221.
Third, Tasks 2,3,5 (Figs.~\ref{fig:result2}, \ref{fig:result_new}, and \ref{fig:result3}) are evaluated by $A_{\mathrm{D1}}*B_{\mathrm{D2}}$, which is for {\it different objects through different diffusers}.  This gives the lowest correlation of around 0.207.

A single-valued metric does not sufficiently capture the rich information encoded in the speckle patterns.  
As inspired by speckle correlography~\cite{labeyrie1970attainment} and the variants~\cite{Bertolotti2012,Katz2014,Edrei2016}, next we investigate the speckle intensity correlation function for different speckle pairs.
Representative examples from our main findings are presented in Fig.~\ref{fig:correlation}(b).
Importantly, taking the speckle intensity autocorrelation as the reference, {\it speckle intensity {\it cross}-correlation from the same object but through two different diffusers (e.g. the first for training, and the second for testing) resembles the similar pattern as the reference. }
These correlation patterns do not follow the simple relation exploited in ~\cite{labeyrie1970attainment,Bertolotti2012,Katz2014,Edrei2016}. 
Nevertheless, the {\it invariance} maintained across speckle patterns {\it from training and testing diffusers} do suggest that there exist {\it learnable} and {\it generalizable} features. 
This suggests that if the CNN is {\it trained and tested with the same object but through different diffusers} (e.g. in Fig.~\ref{fig:result}), there exists physically meaningful invariance exist in these speckle intensity correlation patterns. 
Our CNN model is able to discover and exploit these `hidden' information although these speckle pairs are considered `decorrelated' based on the PCC.
Next, correlation patterns from visually similar objects are shown to present notable difference, which demonstrates the sensitivity of these features. 
Overall, we speculate that these invariant correlation patterns/features could contribute to the scalability  of our CNN with respect to speckle decorrelations.   
Furthermore, our results on {\it unseen objects} through unseen diffusers (Figs.~\ref{fig:result2} and \ref{fig:result_new}) suggest that these learned invariance are generalizable to a broader range of speckle measurements.

\section{conclusion and discussion}

We have demonstrated a deep learning framework to significantly improve the scalability of imaging through scattering.  
Traditional techniques suffer from the {\it `one-to-one'} limitation, in which one model only works for one fixed scattering medium.
Here, we take an entirely different {\it `one-to-all'} strategy, in which one model fits to all scattering media within the same class.  
In practice, this leads to significantly improved resilience to speckle decorrelations and improved space-bandwidth-product.  
Our approach promises  highly scalable, large information-throughput imaging through complex scattering media.

We envision that our technique can be useful in imaging biological samples. 
Several macroscopic parameters~\cite{wang2012biomedical}, such as absorption and scattering coefficients, and (transport) mean-free-path, are routinely used to characterize a sample's  scattering properties, as well as to make phantoms with controlled optical properties. 
One may train, classify, and image through these biological samples by adapting our technique. 

We have demonstrated our technique to image through shift-variant scattering induced by a thin diffuser.  
This condition closely resembles those involving aberrations induced by a single scattering layer~\cite{Ji2010,li2015conjugate}.  
Our technique opens up the opportunity to compensate for these aberrations in real-time without expensive hardware, and provide expanded field-of-views and improved tolerance to the change of aberrations. 
The ultimate challenge for imaging through scattering is to deal with volumetric multiple scattering.  
Several learning-based approaches have been reported recently~\cite{Tian.Waller2015,Waller.Tian2015,kamilov2015learning,Liu.etal2017c,soubies2017efficient,sun2018efficient}. 
Future work could adapt our approach to handle these more challenging  scenarios.

\section*{Funding}
National Science Foundation (NSF) (1711156).

\section*{Acknowledgments}
We thank Xiaojun Cheng for discussions on correlation analysis.

\section*{Conflict of interest}
The authors declare no conflict of interest.



\end{document}